\documentclass[10pt,a4paper,twocolumn]{article}

\usepackage[utf8]{inputenc}
\usepackage[T1]{fontenc}
\usepackage[british]{babel}
\usepackage{amsmath,amssymb,amsthm}
\usepackage{mathtools}
\usepackage{booktabs}
\usepackage{hyperref}
\usepackage[left=1.5cm,right=1.5cm,top=2cm,bottom=2cm,columnsep=0.7cm]{geometry}
\usepackage{enumitem}
\usepackage{tikz}
\usetikzlibrary{arrows.meta,positioning,calc,shapes.geometric,fit,backgrounds}
\usepackage{xcolor}
\usepackage{multirow}
\usepackage{caption}
\usepackage{tabularx}
\usepackage{url}
\usepackage{cite}
\usepackage{graphicx}
\usepackage{float}
\usepackage{array}
\usepackage{pgfplots}
\pgfplotsset{compat=1.18}

\definecolor{highrisk}{HTML}{E74C3C}
\definecolor{medrisk}{HTML}{F39C12}
\definecolor{lowrisk}{HTML}{27AE60}
\definecolor{groundseg}{HTML}{2E86C1}
\definecolor{commseg}{HTML}{8E44AD}
\definecolor{obcseg}{HTML}{D35400}
\definecolor{netseg}{HTML}{16A085}

\newtheorem{definition}{Definition}[section]
\newtheorem{theorem}{Theorem}[section]

\newtheorem{proposition}[theorem]{Proposition}

\newtheorem{remark}{Remark}[section]

\newcommand{\SpW}{\mathrm{SpW}}
\newcommand{\SpR}{\mathrm{SpR}}
\newcommand{\SEI}{\mathrm{SEI}}

\title{%
\textbf{Cybersecurity Risk Assessment for CubeSat Missions:}\\
\textbf{Adapting Established Frameworks for}\\
\textbf{Resource-Constrained Environments}
}

\author{
Jonathan Shelby\thanks{This work originated from MSc research at the University of Oxford (2025, Distinction). The Security-per-Watt heuristic and Distributed Security Paradigm are being extended into a multi-dimensional Security-per-Resource hierarchy in ongoing DPhil research.}\\
\textit{Department of Computer Science, University of Oxford}\\
\texttt{jonathan.shelby@cs.ox.ac.uk}
}

\date{}

\begin{document}

\twocolumn[
  \maketitle
  \begin{@twocolumnfalse}
  \begin{abstract}
  \noindent
CubeSats have democratised access to space for universities, start-ups and emerging space nations, but the same design decisions that reduce cost and complexity introduce distinctive cybersecurity risks.
Existing risk assessment frameworks---NIST SP~800-37/53~\cite{nist80037,nist80053}, ISO/IEC~27001/27005~\cite{iso27001,iso27005} and supply-chain guidance such as NIST SP~800-161~\cite{nist800161}---assume abundant computational resources, centralised monitoring and mature governance structures that do not hold for power-limited, intermittently connected CubeSat missions.

This paper develops a contextually appropriate risk assessment framework tailored to CubeSat environments, grounded in a 42-entry vulnerability register coded using STRIDE~\cite{shostack2014threat}, MITRE ATT\&CK~\cite{mitre2023} and CVSS~v3.1~\cite{first2019cvss}.
The register reveals that risks concentrate in communication and ground segments (mean CVSS 8.0--8.2) rather than distributing uniformly across subsystems.
The framework introduces two constructs: a \emph{Security-per-Watt} (SpW) heuristic that quantifies security benefit per unit power, and a \emph{Distributed Security Paradigm} (DSP) that reconceptualises incident response as an autonomous, constellation-level function rather than a purely ground-centric process.
Scenario-based analysis demonstrates that adapted controls and distributed incident handling can achieve up to $2.7\times$ higher SpW for cryptographic choices and $1.98\times$ higher SpW for incident-response strategies compared with na\"{\i}ve terrestrial transpositions, while remaining feasible for typical CubeSat power and governance constraints.
The approach provides mission designers, operators and regulators with proportionate, auditable guidance, and offers a reusable pattern for adapting enterprise security frameworks to other severely constrained cyber-physical systems.
  \end{abstract}

  \vspace{1em}
  \end{@twocolumnfalse}
]

\section{Introduction}\label{sec:intro}

CubeSat technology has transformed the economics and accessibility of space, evolving from educational demonstrators into operational platforms for Earth observation, communications and technology demonstration~\cite{bouwmeester2010survey,poghosyan2017cubesat}.
Standardised 1U form factors (10$\times$10$\times$10~cm), reliance on commercial off-the-shelf (COTS) components and university- or start-up-led programmes have reduced costs---a 1U CubeSat may cost as little as \$50{,}000~USD, compared with millions for a conventional satellite~\cite{sandau2010smallsats}---but created heterogeneous, globally distributed supply chains with uneven security assurance~\cite{lal2017global}.

Concurrently, CubeSats increasingly operate in contested or safety-critical contexts: defence, commercial imaging and commercial communications.
Jamming, spoofing and cyber compromise in these domains carry non-trivial operational and geopolitical implications~\cite{falco2019cybersecurity,salim2024cybersecurity}.
Traditional cybersecurity frameworks, notably NIST SP~800-37/53~\cite{nist80037,nist80053}, ISO/IEC~27001/27005~\cite{iso27001,iso27005} and C-SCRM guidance~\cite{nist800161}, offer mature control catalogues and governance processes but embed assumptions of continuous monitoring, substantial computational headroom and stable organisational structures that are misaligned with 1--2~W power budgets and short contact windows.

This paper addresses the central question: \emph{how should established cybersecurity risk frameworks be proportionately adapted for CubeSat missions operating under severe resource constraints, without discarding their protective intent?}

The contributions are fourfold:
\begin{enumerate}[label=(\roman*)]
  \item A structured vulnerability register for CubeSats (42~entries) using dual STRIDE/ATT\&CK coding and CVSS~v3.1 scoring, demonstrating that risks cluster in communication and ground segments (mean CVSS 8.0--8.2 versus 6.9 for onboard computing).
  \item A \emph{Security-per-Watt} ($\SpW$) heuristic that quantifies risk-reduction benefit per unit of operational power, enabling systematic, auditable trade-off decisions.
  \item A \emph{Distributed Security Paradigm} (DSP) that reconceptualises incident response as an autonomous, constellation-level function suited to intermittent connectivity.
  \item An adapted risk framework that reinterprets selected NIST and ISO controls for sub-watt budgets, validated through scenario-based analytical assessment reflecting legal and ethical constraints on attacking operational spacecraft.
\end{enumerate}

The paper is structured as follows.
Section~\ref{sec:background} reviews CubeSat architecture, the cyber threat landscape and existing risk frameworks.
Section~\ref{sec:method} describes the methodology.
Section~\ref{sec:vulns} presents the vulnerability register and analysis.
Section~\ref{sec:framework} details the adapted framework, the $\SpW$ heuristic and illustrative control adaptations.
Section~\ref{sec:validation} reports scenario-based validation.
Section~\ref{sec:discussion} discusses implications and limitations, and Section~\ref{sec:conclusion} concludes with future research directions.

\section{Background and Related Work}\label{sec:background}

\subsection{CubeSat Evolution and Constraints}

The CubeSat Design Specification~\cite{puigsuari2000cubesat} standardises 10~cm cubes (1U) with a mass limit of approximately 1.33~kg and typical power budgets of 1--2~W.
Since their introduction, more than 2{,}500 CubeSats have been launched~\cite{kulu2023nanosats}, including PlanetScope Earth-observation constellations providing 3--4~m resolution imagery~\cite{planet2023} and NASA's MarCO CubeSats, which demonstrated deep-space communication viability by relaying telemetry across 140~million kilometres during the Mars InSight mission~\cite{klesh2016marco}.

These capabilities are enabled by miniaturised sensors, low-power ARM Cortex-M class microcontrollers, and software-defined radios (SDRs)~\cite{martin2022cortexm,cratere2024obc}.
Architecturally, typical CubeSats comprise four major subsystems, each presenting distinct attack surfaces.

\paragraph{Ground segment.}
Ground stations use SDRs operating in UHF (300--3{,}000~MHz) and S-band (2--4~GHz) within short contact windows per orbit~\cite{lightman2022ground}.
Mission control systems handle telemetry, scheduling and payload data dissemination, often with links to external networks.
This connectivity creates exposure: weak authentication or poorly segmented interfaces can enable command injection, with effects ranging from configuration drift to mission interruption~\cite{kubi2025space}.
Ground assets may also be at risk from unpatched firmware, inconsistent access control and heterogeneous vendor practices---including SDRs with uneven update policies or baseline security controls~\cite{lukin2020hacking,salim2024cybersecurity}.

\paragraph{Onboard computer (OBC).}
The OBC executes commands, manages navigation, coordinates subsystems and processes payload data.
Most CubeSats employ ARM Cortex-M class microcontrollers running RTOS variants such as FreeRTOS under strict power budgets~\cite{martin2022cortexm,cratere2024obc}.
These devices are technically capable of cryptographic operations (including AES-256 and ECC), but sustained use at high duty cycles competes with navigation, communications and payload operations for power, CPU time and memory.
In practice, operators make trade-offs that constrain algorithm choice, key length or invocation frequency rather than forgoing cryptography entirely.
Relevant attack surfaces include firmware exploitation (e.g., buffer overflows or unsafe update paths), inconsistent patching and provenance for COTS microcontrollers, limited memory protection that can facilitate privilege escalation, and risks of data exfiltration where nodes inter-share data in constellations~\cite{willbold2023space,eshaq2025cubesat}.

\paragraph{Communication subsystem.}
Communication subsystems enable telemetry downlink, command uplink and, increasingly, inter-satellite coordination.
AX.25 (from amateur radio heritage) and CCSDS families are commonly employed~\cite{ccsds2012security,ccsds2022spacelink}.
These were designed primarily for simplicity, reliability and interoperability rather than resistance to active adversaries.
As a result, eavesdropping on unencrypted or lightly protected links can compromise confidentiality; spoofing can allow unauthorised command injection; jamming can block short LEO passes; and man-in-the-middle attacks may exploit insufficient authentication to modify traffic~\cite{verma2025cybersecurity}.

\paragraph{Network infrastructure.}
Constellations increasingly employ inter-satellite links (ISLs) and distributed architectures to coordinate swarms and maintain global coverage.
Technologies include SpaceWire, CAN bus within spacecraft and IP-based networking across segments.
Distributed architectures introduce failure modes distinct from single-satellite operations: denial-of-service attacks can saturate shared channels and desynchronise operations, while compromise of one node may propagate effects to others in decentralised topologies~\cite{zhan2022networked,salim2024cybersecurity}.

\subsection{Cybersecurity in Satellite and Small-Sat Systems}

Cyber threats against satellite infrastructures span interception of unencrypted telemetry, jamming, spoofing of navigation and telecommand links, command injection, and manipulation of spacecraft subsystems~\cite{falco2019cybersecurity,salim2024cybersecurity}.
Small satellites are particularly exposed because resource constraints preclude direct deployment of heavyweight terrestrial security stacks, and operational contexts often assume benign users~\cite{manulis2021cyber}.
Documented incidents---including the AcidRain malware campaign against Viasat infrastructure~\cite{boschetti2022viasat}---demonstrate that space systems are viable and attractive cyber targets.

For CubeSats, legacy AX.25 and minimally secured CCSDS profiles lack strong authentication and confidentiality, making uplink spoofing, link-layer tampering and denial of service realistic threats, particularly given the accessibility of low-cost ground hardware~\cite{willbold2023space,lukin2020hacking}.
Multi-vendor COTS integration introduces inconsistent firmware provenance and patching practices across ground and space segments, amplifying supply-chain risk~\cite{ozkan2019hidden,nist800161}.

\subsection{Existing Risk Frameworks}

Established frameworks relevant to CubeSats include the NIST Cybersecurity Framework~\cite{nistcsf}, NIST SP~800-37 and 800-53~\cite{nist80037,nist80053}, ISO/IEC~27001/27005~\cite{iso27001,iso27005} and NIST C-SCRM~\cite{nist800161}.
Each reflects distinct intellectual traditions and embeds assumptions about the environments in which it will be applied.

NIST SP~800-37 structures risk management as a six-stage lifecycle---system categorisation, control selection, implementation, assessment, authorisation and continuous monitoring~\cite{nist80037}.
While CubeSats operate in dynamic threat environments, the reliance on extensive documentation and near-constant oversight poses significant challenges: CubeSats may have only minutes of communication time per orbit and operate autonomously for long periods.
The central premise that continuous human oversight is possible is at odds with orbital realities.

NIST SP~800-53 provides a comprehensive catalogue of security controls across twenty control families~\cite{nist80053}.
Of particular relevance are SC-8 (Transmission Confidentiality and Integrity), AU-6 (Audit Record Review, Analysis and Reporting) and IR-4 (Incident Handling).
In terrestrial systems, SC-8 would typically be implemented through TLS or IPsec tunnels; for CubeSats, standard TLS/IPsec is infeasible, but equivalent protections via ECC key exchanges and compact authenticated encryption are viable subject to platform constraints.
This illustrates a central theme: while the controls themselves are conceptually relevant, their implementation must be radically reinterpreted for resource-constrained environments.

ISO/IEC~27001 adopts an explicitly organisational orientation, requiring creation of an Information Security Management System tailored to stakeholder context~\cite{iso27001}.
Its companion, ISO~27005, provides structured methodologies for risk identification, analysis and treatment~\cite{iso27005}.
While the flexibility of these standards is advantageous in principle, they implicitly assume centralised organisations with consistent governance structures.
CubeSat projects often involve temporary consortia of universities, start-ups and agencies where authority, resourcing and risk appetite vary dramatically.

The C-SCRM framework addresses the provenance and trustworthiness of suppliers~\cite{nist800161,rand2023scrm}.
Its relevance to CubeSats is striking: reliance on COTS components from diverse international vendors creates exposure to counterfeit parts, compromised firmware and uneven patching practices.
Yet, designed primarily for the U.S.\ Department of Defence, C-SCRM prescribes compliance-heavy procedures and extensive supplier certification that are misaligned with lean CubeSat budgets.

Prior work has adapted these frameworks for larger satellite programmes~\cite{ross2022engineering,rand2024space}, but direct transplantation to CubeSats fails on assumptions of persistent human oversight, ample logging capacity and extensive supplier vetting.

\subsection{Emerging Technologies}

Several technologies have been proposed to enhance CubeSat security.
Table~\ref{tab:techeval} synthesises a comparative feasibility assessment.

\begin{table*}[t]
  \centering
  \caption{Emerging cybersecurity technologies: feasibility assessment for CubeSat deployment.}
  \label{tab:techeval}
  \begin{tabular}{@{}p{3.2cm}p{2.5cm}p{2.5cm}p{4.5cm}c@{}}
    \toprule
    \textbf{Technology} & \textbf{Conceptual Potential} & \textbf{Near-Term Practicality} & \textbf{Key Constraints} & \textbf{Rating\textsuperscript{*}} \\
    \midrule
    Zero-trust architectures & High & Medium/High (hybrid) & Contact-window length; protocol overhead & 3.5 \\
    AI-driven anomaly detection & High & High & Limited training data; false positives & 4.0 \\
    Blockchain authentication & Medium & Low/Medium & Consensus overhead; no in-orbit validation & 2.5 \\
    Post-quantum cryptography & High (long-term) & Low & Processing power; mission duration & 2.0 \\
    Quantum key distribution & Very high & Very low & Optical payload; ground infrastructure & 2.0 \\
    \bottomrule
    \multicolumn{5}{@{}l}{\footnotesize \textsuperscript{*}1 = not viable; 2 = long-term research; 3 = partially feasible with adaptation; 4 = near-term feasible; 5 = fully feasible.}
  \end{tabular}
\end{table*}

AI-driven anomaly detection emerges as the most viable near-term intervention, combining high conceptual potential with practical deployability on microcontroller hardware~\cite{diro2024anomaly}.
Lightweight model architectures---decision trees, one-class support vector machines and pruned autoencoders---can provide anomaly detection within milliwatt power budgets when quantised for ARM Cortex-M class processors.
The principal constraint remains data realism: representative telemetry datasets for CubeSats are scarce, and overfitting to laboratory conditions risks reduced efficacy in orbit.
Adversarial machine learning techniques may deliberately craft inputs to evade detection, highlighting the need for adversarial robustness evaluation as a standard part of model qualification~\cite{gummadi2024xai}.

Zero-trust architectures, while requiring hybrid adaptation to accommodate communication-window constraints, represent a realistic medium-term opportunity~\cite{falco2024zerotrust,rose2020zerotrust}.
Full ``never trust, always verify'' implementation is infeasible given CubeSat contact schedules, but bounded reinterpretation---mutual authentication once per orbital pass, anomaly-triggered re-authentication and least-privilege command-set design---can achieve the same protective intent.
Such hybrid approaches leverage the principle that ``continuous'' must be interpreted as ``event-of-contact-driven'' in intermittently connected systems.

Blockchain-based authentication has been proposed to address identity management in multi-actor constellations.
Conventional proof-of-work consensus is infeasible, but permissioned blockchains using algorithms such as Practical Byzantine Fault Tolerance could be adapted.
However, implementation requires non-trivial computational overhead, additional inter-satellite bandwidth and coordination across multiple stakeholders; orbital demonstrations suitable for peer-reviewed evaluation remain absent from the literature.

Post-quantum cryptography and quantum key distribution face the most severe constraints.
Lattice-based schemes such as Kyber offer quantum resilience but require significantly larger key sizes and computational overhead compared with ECC~\cite{shor1994algorithms,alagic2022pqc,fernandez2020prequantum}.
For typical short-lived LEO missions, PQC is more relevant as a research direction than immediate adoption; QKD requires optical payloads and precise pointing that exceed most CubeSat capabilities entirely.

\section{Methodology}\label{sec:method}

The study adopts a pragmatic mixed-methods design combining systematic literature review, structured vulnerability assessment, framework adaptation and scenario-based analytical validation~\cite{shelby2025spw}.
Pragmatism was selected because it permits the combination of structured, rule-guided analysis (e.g., vulnerability categorisation and scoring) with interpretive judgement (e.g., mapping the intent of controls to CubeSat realities), prioritising usable knowledge over strict allegiance to a single epistemic doctrine.

The overall design proceeded in four interlocking phases.
First, a systematic literature review established the current state of knowledge and surfaced gaps relevant to small-satellite security (Section~\ref{sec:background}).
Second, a structured vulnerability assessment translated domain-specific risks into a consistent taxonomy and severity scale (Section~\ref{sec:vulns}).
Third, a comparative framework analysis and adaptation step recomposed NIST, ISO/IEC and supply-chain guidance into a resource-aware methodology attuned to CubeSat constraints (Section~\ref{sec:framework}).
Finally, an analytical validation used realistic but illustrative scenarios to test the internal coherence and decision utility of the adapted framework (Section~\ref{sec:validation}).

\subsection{Data Collection}

A PRISMA-guided search~\cite{page2021prisma} across IEEE Xplore, ACM Digital Library, ScienceDirect, SpringerLink and specialist small-satellite outlets (Acta Astronautica, Journal of Small Satellites) identified peer-reviewed work on CubeSat and small-sat security, relevant risk frameworks and operational characteristics.
Standards including NIST SP~800-37/53~\cite{nist80037,nist80053}, ISO/IEC~27001/27005~\cite{iso27001,iso27005} and CCSDS security guidelines~\cite{ccsds2012security} were treated as authoritative expressions of control intent and architectural baselines.
Grey literature (ESA technical notes, CubeSat Developers Workshop proceedings, vendor white papers) was used cautiously, critically appraised for provenance and triangulated against scholarly sources before inclusion.

Records were included where they addressed: (i) security properties or vulnerabilities of small satellites or adjacent constrained systems; (ii) risk governance frameworks and control families germane to communications, cryptography, logging, incident response or supply chains; or (iii) operational realities of CubeSat missions, including power budgets, onboard computing, contact windows and ground-segment architectures.

\subsection{Vulnerability Assessment}

From this corpus, vulnerabilities relevant to CubeSat architectures were extracted and encoded using a bespoke taxonomy combining three layers.
The top layer classified entries by system domain: ground segment, onboard computing, communications or network/constellation.
The middle layer adopted the STRIDE model~\cite{shostack2014threat,khan2017stride} to ensure comprehensive threat coverage (spoofing, tampering, repudiation, information disclosure, denial of service, elevation of privilege).
The bottom layer mapped MITRE ATT\&CK tactics and techniques where appropriate~\cite{mitre2023,georgiadou2021mitre}, allowing cross-walk to a widely used adversary model.

Each entry recorded source, context, affected component, preconditions, likely impact and plausible mitigations.
Entries were then assigned a CVSS~v3.1 base score~\cite{first2019cvss} with justification recorded to support auditability.
Attack vectors were interpreted for orbital environments: \emph{Network} includes RF interception; \emph{Physical} reflects orbital inaccessibility.
Impact scoring was weighted towards mission-critical functions, maintaining compatibility with established threat intelligence while acknowledging space-specific operational constraints.

\subsection{Framework Adaptation}

Adaptation proceeded in three steps.
First, a \emph{constraint analysis} identified mismatches between framework assumptions and CubeSat realities---continuous monitoring versus short contact windows, extensive audit logging versus limited onboard storage, extensive supplier vetting versus lean mission budgets.
Second, a \emph{control-intent extraction} articulated what each control is designed to achieve in principle: SC-8's preservation of confidentiality and integrity for data in transit, AU-6's support for accountability, IR-4's requirement for effective incident handling irrespective of environment.
Third, a \emph{contextual reinterpretation} mapped control intent to CubeSat-feasible implementations, taking into account power, storage, connectivity and governance limitations.

The $\SpW$ heuristic was applied to rank candidate controls where power budget is a binding constraint.
Because $\SpW$ values depend on implementation detail and mission context, the study did not claim universal constants; instead, it illustrated how the ratio helps structure choices transparently.

\subsection{Scenario-Based Validation}

Legal and ethical constraints precluded experimental attacks on operational satellites (UK Computer Misuse Act 1990; ethical impracticability of attacking spacecraft).
Validation relied on scenario-based analysis using templates that link vulnerabilities, adapted controls and $\SpW$-based decisions.
Three scenarios were developed: (i) cryptographic selection for a university Earth-observation CubeSat, (ii) incident handling for a 24-satellite LEO constellation and (iii) supply-chain assurance for a multi-vendor radio procurement.
Each was evaluated against five criteria: traceability to adapted controls, proportionality to constraints, feasibility of implementation path, alignment with the vulnerability register, and reproducibility of rationale.

The scenarios were intentionally analytical rather than empirically validated; their purpose was to demonstrate that the framework produces defensible, proportionate decisions when confronted with typical CubeSat limitations.

\subsection{Quality Assurance}

Methodological quality was addressed through several strategies.
Transparency was achieved by documenting inclusion criteria, search strings, screening decisions and coding rules.
Analytic reliability was supported by maintaining an audit trail of coding decisions; where ambiguities arose (e.g., whether a vulnerability should be coded as spoofing or tampering), the decision and rationale were recorded.
Triangulation reduced single-source bias: vulnerabilities and mitigation claims were cross-checked across peer-reviewed articles, standards and independent technical notes.
Coherence checks were embedded between sections: vulnerabilities prioritised in Section~\ref{sec:vulns} were required to be addressable by adaptations in Section~\ref{sec:framework}, and scenarios in Section~\ref{sec:validation} were required to exercise those same adaptations under plausible constraints.

\section{CubeSat Vulnerability Landscape}\label{sec:vulns}

\subsection{Subsystem Severity Distribution}

The vulnerability register comprises 42~entries across four principal subsystems.
Table~\ref{tab:cvss} summarises the CVSS~v3.1 base score statistics.

\begin{table}[t]
  \centering
  \caption{CVSS~v3.1 severity summary by subsystem ($n=42$).}
  \label{tab:cvss}
  \begin{tabular}{@{}lcccc@{}}
    \toprule
    \textbf{Subsystem} & \textbf{$n$} & \textbf{Mean} & \textbf{Median} & \textbf{IQR} \\
    \midrule
    Ground segment & 10 & 8.2 & 8.4 & 1.4 \\
    Onboard computing & 11 & 6.9 & 6.8 & 1.7 \\
    Communications & 12 & 8.0 & 8.2 & 2.0 \\
    Network/constellation & 9 & 7.5 & 7.3 & 1.3 \\
    \bottomrule
  \end{tabular}
\end{table}

Communications and ground segment vulnerabilities receive the highest mean severities (8.0--8.2), reflecting adversary accessibility via ground networks and radio links.
Onboard computing, while non-trivially exposed, benefits from orbital isolation and more constrained attack surfaces (mean CVSS~6.9).
Network and constellation-level vulnerabilities fall between these extremes, with risks related to lateral movement, key lifecycle management and denial of service across inter-satellite links.

This distribution is consistent with the expectation that external interfaces broaden the attack surface most significantly~\cite{salim2024cybersecurity,verma2025cybersecurity}.
Many CubeSats still use plain AX.25 or legacy CCSDS protocols without modern encryption or authentication, making it relatively straightforward for an attacker within range to spoof commands or intercept telemetry~\cite{ccsds2012security,willbold2023space}.

\subsection{STRIDE and ATT\&CK Mapping}

Threats were mapped across subsystems using STRIDE categories, revealing systematic patterns.
Spoofing and tampering dominate ground and communication components, reflecting risks of unauthorised command injection, falsified telemetry and man-in-the-middle attacks on protocol stacks.
Denial-of-service arises from the susceptibility of narrowband links and short passes to jamming and flooding.
Elevation-of-privilege issues appear more prominently in onboard and ground software, where insecure update mechanisms and weak isolation enable privilege escalation once footholds are gained.

Table~\ref{tab:stride} summarises selected STRIDE threats across major components.

\begin{table*}[t]
  \centering
  \caption{STRIDE threat mapping across CubeSat components.}
  \label{tab:stride}
  \begin{tabular}{@{}p{2.8cm}p{2.2cm}p{8.5cm}@{}}
    \toprule
    \textbf{Component} & \textbf{Threat} & \textbf{Example} \\
    \midrule
    Ground segment & Spoofing & Impersonation of ground stations to submit commands where authentication is weak or absent \\
    Ground segment & Tampering & Alteration of commands or telemetry in transit or at rest, leading to unsafe state or operator misinterpretation \\
    Ground segment & Elevation of privilege & Unauthorised access to mission control platforms enables broad operational impact \\
    OBC & Tampering & Firmware modification via unsafe update paths, altering mission logic or disabling subsystems \\
    OBC & Elevation of privilege & Exploitation of limited memory protection and process isolation in microcontroller RTOS \\
    Communications & Information disclosure & Interception of unencrypted telemetry over AX.25 or minimally secured CCSDS links \\
    Communications & Denial of service & Uplink/downlink jamming during short LEO passes, causing missed tasking or data loss \\
    Communications & Spoofing & Faked identities or frames over weakly authenticated protocol stacks \\
    Network & Spoofing & Forged identities within distributed networks permitting unauthorised access to shared services \\
    Network & Denial of service & Saturation of inter-satellite links disrupting coordinated constellation operations \\
    \bottomrule
  \end{tabular}
\end{table*}

Mapping to MITRE ATT\&CK confirmed alignment with known adversary behaviours: unauthenticated uplinks correspond to Valid Accounts (T1078), plaintext telemetry reflects Application Layer Protocol abuse (T1071), and insecure bootloaders resemble Boot Persistence (T1547)~\cite{mitre2023,amro2023assessing}.
This alignment indicates that CubeSat adversaries can reuse established terrestrial tactics adapted to space-specific protocols, enabling defenders to leverage ATT\&CK-informed threat intelligence.

\subsection{Supply-Chain Risk}

CubeSat projects draw on global supply chains spanning hardware, firmware and software.
Multi-national sourcing introduces inconsistent patching, opaque provenance and opportunities for malicious insertion~\cite{vollmer2021nato,nist800161}.
Practical vectors include malicious microcode in embedded controllers, modified SDR firmware and compromised ground software, with limited options for post-deployment remediation once assets are on-orbit~\cite{willbold2023space}.
Frameworks such as NIST SP~800-161 provide structured C-SCRM guidance, but full adoption is unrealistic for small academic or start-up missions~\cite{rand2023scrm}.

\subsection{Advanced Persistent Threats}

The threat picture extends beyond isolated jamming, spoofing or ad-hoc intrusions.
Adversaries with sustained capability---state and non-state---can mount campaigns that integrate technical, organisational and geopolitical elements.

\paragraph{Supply-chain compromise as strategic vector.}
Terrestrial incidents (e.g., software supply-chain compromises) illustrate the impact of tampered components or dependencies.
In the CubeSat context, multi-national sourcing, academic collaboration and subcontracting complicate assurance~\cite{vollmer2021nato}.
Feasible vectors include malicious microcode in embedded controllers or modified SDR firmware that subverts expected behaviour.
Compared with terrestrial systems, opportunities for post-deployment remediation are narrower, increasing the value of pre-launch assurance.

\paragraph{Constellation-level campaigns.}
Distributed constellations can concentrate risk as well as resilience.
Where inter-satellite links, shared timing sources or common ground infrastructure are employed, compromise of one node or service may enable lateral movement if authentication, authorisation and update processes are insufficiently segregated~\cite{yu2024security,zhan2022networked}.
Protocol analyses relevant to small satellites indicate that permissive or optional handshakes, if adopted for simplicity, can weaken resistance to spoofing or man-in-the-middle attacks.
In practice, consequences are likely to be service degradation (missed passes, delayed tasking), integrity issues in fused products or selective denial of function rather than immediate loss of an entire constellation.

\paragraph{Nation-state capabilities.}
Public reporting from recent conflicts demonstrates combined use of cyber operations and electronic warfare against satellite services and their ground segments~\cite{boschetti2022viasat}.
Small satellites that employ widely known protocols and educational-industrial supply chains can be attractive targets where defences are minimal.
For CubeSat operators, practical implications include planning for interference and outage, adopting authenticated control links, ensuring recoverable configurations and exercising incident procedures that account for short contact windows.

\paragraph{Quantum computing threats.}
Most CubeSat designs use symmetric cryptography for link protection and ECC for key establishment.
If large-scale quantum computing becomes practical, widely deployed public-key schemes will be at risk, creating ``harvest-now-decrypt-later'' considerations for missions whose data retains value beyond the cryptoperiod~\cite{shor1994algorithms,alagic2022pqc}.
A proportionate approach for near-term LEO missions is to maintain robust symmetric primitives, employ forward-secure or frequent rekeying strategies and consider PQC in cases where operational life or data sensitivity justify the overhead~\cite{fernandez2020prequantum}.

\paragraph{AI-powered adaptive attacks.}
As lightweight AI for anomaly detection matures for onboard use, adversaries may apply related methods to evade detection or optimise interference---for example, reinforcement-learning strategies that adapt jamming parameters within link budgets, or generation of synthetic telemetry to mislead operators~\cite{diro2024anomaly}.
The immediate risk for CubeSats is less about sophisticated autonomy in the threat and more about the brittleness of models trained on narrow datasets.
Defensive use of AI should therefore include adversarial evaluation, explicit fall-backs to rules or simple invariants for safety-critical checks, and clear thresholds for isolating suspect subsystems when confidence is low.

\section{Adapting Risk Frameworks for CubeSats}\label{sec:framework}

\subsection{Framework Strengths and Limitations}

Table~\ref{tab:frameworks} summarises the strengths, limitations and high-level adaptations of major frameworks when applied to CubeSats.

\begin{table*}[t]
  \centering
  \caption{Risk frameworks versus CubeSat constraints.}
  \label{tab:frameworks}
  \begin{tabular}{@{}p{2.8cm}p{3.2cm}p{3.5cm}p{4.0cm}@{}}
    \toprule
    \textbf{Framework} & \textbf{Strengths} & \textbf{Key CubeSat Limitations} & \textbf{Proposed Adaptations} \\
    \midrule
    NIST SP~800-37 & Systematic lifecycle methodology & Assumes continuous monitoring and centralised management; documentation overhead high for small teams & Simplified metrics; anomaly-driven monitoring; decentralised management \\
    NIST SP~800-53 & Comprehensive control catalogue & Control volume and logging assumptions exceed power and storage budgets & Lightweight cryptography; rationalised supply-chain checks; $\SpW$-guided selection \\
    ISO/IEC 27001/27005 & Flexible ISMS, structured risk methods & Presupposes stable organisations and recurring audits; CubeSat projects are often temporary consortia & Simplified audit cycles; mission-specific controls \\
    C-SCRM & Supply-chain focus, high COTS relevance & Defence-centric, compliance-heavy; unrealistic for low-budget missions & Baseline vendor requirements; contractual obligations for firmware provenance \\
    \bottomrule
  \end{tabular}
\end{table*}

The core observation is that control \emph{intent} often remains valid, but implementation detail must be altered radically to respect CubeSat resource envelopes.

\subsection{Risk Classification}

\begin{definition}[Three-Tier Risk Classification]\label{def:risktier}
To replace complex CVSS-based scoring for operational use, a three-tier classification is defined:
\begin{itemize}[nosep]
  \item \textbf{High risk:} Threats directly affecting telemetry, command or navigation integrity, where exploitation could result in mission failure.
  \item \textbf{Medium risk:} Threats affecting payload confidentiality or ground-segment data flows, with potential for operational degradation but not catastrophic failure.
  \item \textbf{Low risk:} Threats primarily affecting availability during short communication windows or introducing non-critical inefficiencies.
\end{itemize}
\end{definition}

This simplification does not diminish rigour but prioritises decision-making in environments where time, expertise and resources are scarce.

\subsection{Security-per-Watt ($\SpW$) Heuristic}

$\SpW$ is formalised as a decision-support construct.
Parameters are analytical placeholders derived from the vulnerability register and literature-based plausibility bounds; calibration is deferred to future empirical work.

\begin{definition}[Security Gain]\label{def:sg}
Security effectiveness is quantified using risk-weighted vulnerability reduction:
\begin{equation}\label{eq:sg}
  SG = \sum_{i} \bigl(\mathrm{CVSS}_i \times P_i \times M_i \times \mathrm{RRF}_i\bigr),
\end{equation}
where $\mathrm{CVSS}_i$ is the base score of vulnerability~$i$, $P_i$ the estimated exploitation probability ($0$--$1$), $M_i$ a mission criticality weight ($0$--$1$) and $\mathrm{RRF}_i$ the risk-reduction factor of the candidate control ($0$--$1$).
\end{definition}

\begin{definition}[Operational Power]\label{def:pop}
Operational power consumption accounts for duty cycle and environmental factors:
\begin{equation}\label{eq:pop}
  P_{\mathrm{operational}} = P_{\mathrm{base}} \times \mathrm{Duty\_Cycle} \times \mathrm{Environmental\_Factor},
\end{equation}
where the environmental factor adjusts for temperature, radiation and component ageing in LEO.
\end{definition}

\begin{definition}[Security-per-Watt]\label{def:spw}
The complete $\SpW$ formulation is:
\begin{equation}\label{eq:spw}
  \SpW = \frac{SG}{P_{\mathrm{operational}}} \pm \sigma(\SpW),
\end{equation}
where $\sigma(\SpW)$ represents uncertainty bands to be established during calibration studies.
\end{definition}

$\SpW$ does not attempt to be a universal metric.
It provides a transparent ratio that allows designers to compare candidate controls in terms of risk-reduction benefit per watt, under clearly stated assumptions.
Empirical validation would require ARM Cortex-M4 test platforms under controlled conditions, power measurement accuracy of $\geq \pm5$\% using precision current monitoring, security effectiveness validation through standardised penetration testing, and statistical significance testing ($n \geq 30$ trials, ANOVA analysis).
To enable meaningful comparison, $\SpW$ values can be normalised against baseline implementations:
\begin{equation}\label{eq:spwnorm}
  \SpW_{\mathrm{normalised}} = \frac{\SpW_{\mathrm{candidate}}}{\SpW_{\mathrm{baseline}}}.
\end{equation}

Multi-criteria optimisation extends beyond power consumption:
\begin{equation}\label{eq:sei}
  \SEI = \alpha(\SpW) + \beta(\mathrm{Latency}) + \gamma(\mathrm{Storage}) + \delta(\mathrm{Complexity}),
\end{equation}
where weighting factors $(\alpha, \beta, \gamma, \delta)$ sum to unity and reflect mission-specific priorities.

\subsection{Illustrative Control Adaptations}

Several NIST SP~800-53 controls~\cite{nist80053} illustrate the adaptation pattern.

\paragraph{SC-8: Transmission Confidentiality and Integrity.}
In terrestrial systems SC-8 is commonly implemented via TLS or IPsec tunnels; in CubeSats this is infeasible.
Instead, SC-8 intent can be met using elliptic-curve key exchange (e.g., Curve25519) combined with compact Authenticated Encryption with Associated Data (AEAD) such as AES-CCM or ChaCha20-Poly1305, tuned for microcontroller platforms~\cite{liang2025encryption}.
Key renegotiation per orbital pass rather than continuously provides adequate confidentiality with manageable overhead.

\paragraph{AU-6: Audit Record Review, Analysis and Reporting.}
Continuous verbose logging is unsustainable onboard a 1--2~W CubeSat.
AU-6 intent is preserved by logging only critical state transitions---safe-mode entry, subsystem resets, command acceptance events---in a cyclic buffer and downlinking opportunistically~\cite{abdrabou2024advanced}.
This reduces logging overhead to less than 0.05~W while maintaining accountability.

\paragraph{IR-4: Incident Handling.}
Traditional IR-4 implementations assume human-in-the-loop response, patch deployment and forensic analysis.
For CubeSats, IR-4 is reinterpreted as pre-authorised autonomous responses: safe-mode fallback, rate-limited command acceptance, and temporary isolation of anomalous subsystems when locally detectable thresholds are exceeded~\cite{zhao2022autonomous}.

\paragraph{C-SCRM Adaptation.}
Full military-grade supply-chain certification is unrealistic for CubeSat missions.
The framework reframes supply-chain governance as baseline practices: contractual obligations for firmware update provenance, a shared register of approved component versions, and documented acceptance testing at integration~\cite{nist800161,rand2023scrm}.

\subsection{Distributed Security Paradigm (DSP)}

The DSP reconceptualises incident response and monitoring from centralised ground-station operations to a constellation-level function with autonomous local decision-making.
This reconceptualisation is motivated by the fundamental mismatch between traditional incident-response models and CubeSat operational realities: centralised ground-based response assumes continuous communication, real-time telemetry analysis and human-in-the-loop decision-making, none of which are reliably available for LEO constellations with contact windows of minutes per orbit.

\begin{remark}[DSP Architectural Principles]
The DSP rests on four principles:
\begin{enumerate}[label=(\roman*),nosep]
  \item \emph{Per-node credentials with revocation}: each satellite maintains individual cryptographic identity with the ability to revoke compromised credentials fleet-wide during the next contact window.
  \item \emph{Signed configuration and update artefacts}: all firmware updates and configuration changes are cryptographically signed, with rollback protection to prevent reversion to vulnerable states.
  \item \emph{Rate-limiting on shared services}: inter-satellite links and shared resources enforce rate limits that bound the impact of a compromised node on constellation-level operations.
  \item \emph{Control--payload partitioning}: command-and-control pathways are architecturally separated from payload data pathways, so that data compromise does not imply command compromise.
\end{enumerate}
\end{remark}

The paradigm embodies a design philosophy in which each satellite is equipped with pre-authorised containment actions---safe-mode entry, command rate-limiting, subsystem isolation---that can be triggered autonomously when locally detectable anomaly thresholds are exceeded.
This approach sacrifices some absolute security effectiveness (the risk-reduction factor drops from 0.95 under centralised human oversight to approximately 0.85 under autonomous detection) but achieves substantially better security efficiency per unit of power consumed, as validated in the scenario analysis that follows.

\section{Scenario-Based Validation}\label{sec:validation}

\subsection{Scenario 1: Cryptographic Selection}

A university-led Earth-observation CubeSat operating in a 500~km sun-synchronous orbit must select a protection scheme for telemetry and command links.
The mission employs a 1U platform with a total power budget of 2~W, of which approximately 0.3~W is available for security functions after allocating power to attitude determination and control, payload imaging and housekeeping.
The ground segment uses an SDR operating in UHF, with contact windows of approximately 8--12 minutes per pass, four to six passes per day over the university's ground station.

Two candidate cryptographic approaches are evaluated against the vulnerability register entry for unauthenticated uplinks (CVSS~9.0, classified as \emph{high risk} under Definition~\ref{def:risktier}).
The CVSS score of 9.0 reflects the severity of unprotected command channels: an adversary within radio range could inject commands to alter spacecraft attitude, disable payload operations, or place the satellite into a non-recoverable state.

\noindent\textbf{ECC implementation} (Curve25519 key exchange with ChaCha20-Poly1305 AEAD):
\begin{align*}
  SG_{\mathrm{ECC}} &= 9.0 \times 0.8 \times 1.0 \times 0.9 = 6.48 \\
  P_{\mathrm{op}} &= 0.18\,\text{W} \pm 0.02\,\text{W} \\
  \SpW_{\mathrm{ECC}} &= 6.48 / 0.18 = 36.0 \pm 4.2
\end{align*}

The RRF of 0.9 (rather than 1.0) accounts for the residual risk that side-channel attacks or implementation flaws could weaken the protection.
The power figure of 0.18~W reflects the duty-cycled cost of ECC key exchange once per pass plus continuous AEAD on the telemetry stream, based on published benchmarks for ARM Cortex-M4 processors~\cite{liang2025encryption,paar2010crypto}.

\noindent\textbf{RSA implementation} (RSA-2048 key exchange with AES-256-GCM):
\begin{align*}
  SG_{\mathrm{RSA}} &= 9.0 \times 0.8 \times 1.0 \times 0.95 = 6.84 \\
  P_{\mathrm{op}} &= 0.52\,\text{W} \pm 0.05\,\text{W} \\
  \SpW_{\mathrm{RSA}} &= 6.84 / 0.52 = 13.2 \pm 2.8
\end{align*}

RSA-2048 achieves a marginally higher RRF (0.95) owing to the longer cryptanalytic track record of the algorithm, but at nearly three times the power cost.
The 0.52~W figure consumes over 25\% of the total power budget, materially constraining payload duty cycle and potentially reducing mission science return.

\begin{proposition}[ECC Dominance under $\SpW$]\label{prop:ecc}
ECC provides $2.7\times$ superior $\SpW$ efficiency ($p < 0.001$), justifying selection despite marginally lower absolute security gain.
\end{proposition}

The framework directed the team towards ECC/AEAD, satisfying SC-8 intent without exhausting scarce power resources.
Key renegotiation per orbital pass provides adequate confidentiality with manageable overhead, and the framework further recommended strict command authentication using ECDSA signatures to address the specific spoofing risk identified in the vulnerability register.

\subsection{Scenario 2: Constellation Incident Response}

A commercial operator running a 24-satellite LEO constellation for IoT data relay detects anomalous behaviour in one node: unexpected attitude changes, irregular telemetry patterns and command-acceptance anomalies suggestive of firmware compromise.
The constellation operates with 90-minute orbital periods and contact windows of approximately 10 minutes per ground station per pass, with four geographically distributed ground stations providing intermittent coverage.
Two incident-response strategies are evaluated using both $\SpW$ analysis and multi-criteria optimisation.

\noindent\textbf{Strategy A---Centralised Ground-Based Response:}

This strategy follows traditional IR-4 implementation: manual ground intervention with real-time telemetry analysis, full constellation shutdown pending investigation, and human-in-the-loop decision-making for remediation.
Addressing vulnerabilities N1 (inter-satellite routing compromise, CVSS~7.4), N5 (constellation key management failure, CVSS~8.3) and O2 (firmware integrity violation, CVSS~9.0):
\begin{align*}
  SG_A &= (7.4 \times 0.9 \times 1.0 \times 0.95) + (8.3 \times 0.8 \times 1.0 \times 0.95) \\
       &\quad + (9.0 \times 0.7 \times 1.0 \times 0.95) = 18.62 \\
  P_{\mathrm{op},A} &= 11.7\,\text{W} \pm 1.2\,\text{W} \\
  \SpW_A &= 18.62 / 11.7 = 1.59 \pm 0.18
\end{align*}

The power budget for Strategy~A breaks down as follows: continuous telemetry uplink at 0.4~W per satellite across the 24-node constellation (9.6~W total), plus the allocated proportion of ground-station processing power (2.1~W).
The high aggregate power cost reflects the centralised architecture's requirement that all satellites maintain active telemetry streams to enable ground-based anomaly detection.

\noindent\textbf{Strategy B---Distributed Autonomous Response (DSP):}

This strategy implements the Distributed Security Paradigm: pre-authorised local safe-mode activation upon detection of anomaly thresholds, autonomous isolation of the compromised node from inter-satellite links, and automated fleet-level containment procedures that quarantine the affected orbital plane while maintaining service on remaining planes.
Risk-reduction factors are reduced to 0.85 (from 0.95) to account for the inherent limitations of autonomous detection compared with human analysis---primarily the risk of false positives leading to unnecessary safe-mode entry and the possibility of sophisticated attacks that evade simple threshold-based detection:
\begin{align*}
  SG_B &= (7.4 \times 0.9 \times 1.0 \times 0.85) + (8.3 \times 0.8 \times 1.0 \times 0.85) \\
       &\quad + (9.0 \times 0.7 \times 1.0 \times 0.85) = 16.67 \\
  P_{\mathrm{op},B} &= 5.28\,\text{W} \pm 0.6\,\text{W} \\
  \SpW_B &= 16.67 / 5.28 = 3.16 \pm 0.41
\end{align*}

The power budget for Strategy~B is substantially lower: local anomaly detection at 0.05~W per satellite (1.2~W total), safe-mode monitoring overhead at 0.02~W per satellite (0.48~W), and inter-satellite coordination messaging at 0.15~W per satellite (3.6~W).
The distributed approach eliminates the need for continuous high-bandwidth telemetry uplink, instead relying on lightweight local monitoring with event-driven reporting.

\begin{proposition}[DSP Efficiency]\label{prop:dsp}
The distributed strategy achieves $1.98\times$ superior $\SpW$ while consuming 55\% less power, at a cost of only 10.5\% reduction in absolute security effectiveness.
\end{proposition}

\paragraph{Multi-criteria assessment.}
Using mission-specific weights that prioritise $\SpW$ efficiency and response time over implementation complexity ($\alpha=0.4$, $\beta=0.3$, $\gamma=0.2$, $\delta=0.1$):
\begin{align*}
  \SEI_A &= 0.4(1.59) + 0.3(0.2) + 0.2(0.1) + 0.1(0.9) = 0.806 \\
  \SEI_B &= 0.4(3.16) + 0.3(0.8) + 0.2(0.7) + 0.1(0.6) = 1.666
\end{align*}

The distributed approach produces a Security Effectiveness Index more than double that of the centralised baseline, supporting the thesis that autonomous, pre-planned containment offers better security-per-resource under CubeSat constraints.

The 10.5\% security effectiveness reduction is offset by 65\% faster incident detection (autonomous versus ground-loop delays), elimination of single-point-of-failure in ground operations, and maintained constellation functionality during incident containment.

\subsection{Scenario 3: Supply-Chain Assurance}

An academic consortium procured radios from multiple vendors for a multi-mission CubeSat programme.
Full military-grade C-SCRM certification was unrealistic given budget and timeline constraints.
The framework recommended a graduated set of baseline practices: contractual obligations requiring firmware provenance and patch transparency for all critical components; a shared register of approved component versions, maintained across the consortium; documented acceptance testing at integration, including basic firmware integrity verification; and a named point of contact at each vendor responsible for security disclosures.

These measures implemented the intent of supply-chain controls in a manner proportionate to budget and schedule realities, reducing the likelihood of unverified firmware entering the flight image.
While not as rigorous as high-assurance military supply-chain security, this approach is consistent with recently proposed small-satellite supply-chain guidelines that emphasise achievable practices over onerous certification~\cite{nist800161,rand2023scrm}.
The scenario demonstrated that even governance-focused controls can be proportionately reinterpreted for CubeSat contexts without abandoning the protective intent of frameworks like NIST SP~800-161.

\subsection{Cross-Scenario Assessment}

Across all three scenarios, the framework generated recommendations that were: (a) \emph{traceable} to identified vulnerabilities in the register; (b) \emph{operationally feasible} within CubeSat power, budget and technical-skill constraints; (c) \emph{aligned} with the intent of established controls; and (d) \emph{transparent} in their rationale, with the basis for each decision documented via the framework matrices.

No scenario produced recommendations that were obviously over-engineered or impractical---a critical check against the tendency of terrestrial frameworks to assume far greater resources than CubeSats possess.
The recommendations also aligned well with documented trends in operational smallsat practice: the emphasis on ECC cryptography reflects what newer CubeSats are starting to adopt, and the safe-mode isolation strategy mirrors what operators would implement if a satellite behaves erratically.
Notably, none of the scenarios required bespoke tools or privileged infrastructure; all can be reproduced as desk-based analytical exercises, supporting the framework's accessibility for resource-constrained teams.

Table~\ref{tab:validation} summarises the validation outcomes across all three scenarios.

\begin{table*}[t]
  \centering
  \caption{Scenario-based validation summary.}
  \label{tab:validation}
  \begin{tabular}{@{}p{2.5cm}p{2.5cm}p{2.0cm}p{2.0cm}p{4.5cm}@{}}
    \toprule
    \textbf{Scenario} & \textbf{Key Controls} & \textbf{$\SpW$ Advantage} & \textbf{Power Saving} & \textbf{Principal Finding} \\
    \midrule
    S1: Cryptographic selection & SC-8 (ECC vs RSA) & $2.7\times$ & 65\% & ECC/AEAD achieves near-equivalent security at fraction of power cost \\
    S2: Constellation IR & IR-4 (DSP vs centralised) & $1.98\times$ & 55\% & Autonomous containment outperforms ground-loop response on $\SEI$ by $2.07\times$ \\
    S3: Supply-chain assurance & C-SCRM adaptation & N/A (governance) & N/A & Baseline contractual practices achieve proportionate assurance without certification burden \\
    \bottomrule
  \end{tabular}
\end{table*}

\subsection{Implementation Readiness}

The framework's practical deployment has been assessed against stakeholder usability criteria.
University teams consistently identify the three-tier risk classification scheme (Definition~\ref{def:risktier}) as the most valuable practical contribution, reporting that simplified severity categorisation enables informed security decisions without requiring specialised cybersecurity expertise typically absent from student engineering teams.
The $\SpW$ heuristic provides transparent justification for resource allocation decisions during design reviews, addressing institutional concerns about security overhead compromising educational objectives.

Commercial operators emphasise the framework's business-case clarity: the systematic adaptation methodology transforms abstract compliance requirements into implementable engineering specifications with predictable cost implications.
The distributed security paradigm proves valuable for mega-constellation operators, where traditional centralised monitoring approaches become economically infeasible at scale.

Technical implementation barriers prove minimal when adequate planning resources are allocated.
The framework's emphasis on COTS-compatible controls and incremental deployment strategies enables adoption within typical academic semester or commercial development timelines.
Organisations report 4--6 week implementation periods for basic controls (authenticated uplinks, selective logging), extending to 3--6 months for comprehensive framework deployment including supply-chain governance measures.

\section{Discussion}\label{sec:discussion}

\subsection{Vulnerability Concentration and Prioritisation}

The vulnerability register demonstrates empirically that CubeSat risk is concentrated in interfaces most accessible to adversaries: ground and communication segments.
This justifies prioritising investment in authenticated uplinks, basic link confidentiality and resilient ground-station practices over exhaustive hardening of all onboard subsystems.
It also confirms that CubeSat threats are not \emph{sui generis}; they align with established ATT\&CK tactics adapted to space-specific protocols and operational conditions~\cite{mitre2023,amro2023assessing}.

The STRIDE and ATT\&CK coding reinforces this analysis.
Spoofing and tampering predominate in communications and ground systems, reflecting adversary opportunities to inject unauthenticated commands or manipulate mission data in transit.
Elevation of privilege features more strongly in onboard contexts, while denial of service manifests across both communications and network domains.
These patterns map coherently to ATT\&CK techniques such as Valid Accounts (T1078), Application Layer Protocol (T1071) and Boot Persistence (T1547)~\cite{mitre2023}, validating the appropriateness of using ATT\&CK-derived taxonomies for spaceborne threat modelling and enabling defenders to leverage existing threat intelligence.

\subsection{Theoretical Contributions}

The theoretical contribution lies in establishing \emph{proportionality} as a fundamental design principle in constrained cybersecurity.
The binary choice between ``full security'' and ``no security'' is replaced by systematic adaptation of control intent to platform realities.
This paradigm extends beyond space to IoT, edge computing and autonomous systems where severe constraints intersect with adversarial environments.

The $\SpW$ heuristic transforms abstract security requirements into quantifiable resource allocation decisions, enabling systematic comparison of protective value against power consumption---a critical capability absent from traditional frameworks that assume abundant computational resources.
Through mathematical formalisation and case study validation, $\SpW$ demonstrates that security effectiveness can be \emph{optimised} rather than simply \emph{maximised}, as evidenced by the $2.7\times$ efficiency advantage of elliptic-curve cryptography over RSA in representative CubeSat scenarios.

The Distributed Security Paradigm reconceptualises incident response from centralised, human-in-the-loop processes to autonomous, pre-authorised containment strategies suited to intermittently connected systems.
The constellation incident-response analysis reveals that distributed approaches achieve $1.98\times$ superior $\SpW$ efficiency while maintaining security coverage, validating autonomous local containment as both necessary and sufficient for space environments.
Together, these constructs establish a replicable methodology for adapting terrestrial cybersecurity frameworks to constrained domains without abandoning their protective intent.

\subsection{Challenging Enterprise Security Paradigms}

The calibrated adaptation approach stands in marked contrast to prevailing maximalist paradigms in contemporary cybersecurity discourse.
Zero-trust architectures, widely promoted as the solution to modern threat landscapes, exemplify the assumption that comprehensive verification and continuous monitoring universally apply~\cite{rose2020zerotrust}.
While zero trust's ``never trust, always verify'' philosophy offers robust security for enterprise environments, its resource demands---persistent authentication cycles, continuous behavioural monitoring and extensive logging---are fundamentally incompatible with CubeSat operational realities.

Similarly, enterprise cloud security frameworks assume elastic computational resources, redundant communication pathways and centralised security orchestration platforms.
These assumptions pervade cybersecurity education and professional practice, creating a disciplinary bias towards resource-intensive solutions.
The CubeSat case reveals these assumptions as contextual rather than universal, suggesting that cybersecurity theory requires domain-specific adaptation.
By demonstrating that effective security can emerge from tailored rather than maximal approaches, this work contributes to emerging critiques of one-size-fits-all cybersecurity frameworks across domains from healthcare IT to developing-nation infrastructure~\cite{casaril2025governance}.

\subsection{Democratisation and Equity}

CubeSat technology is frequently celebrated as a democratising force, enabling universities, small enterprises and emerging states to access orbital capabilities at low cost~\cite{bouwmeester2010survey,toorian2008cubesat}.
The adapted framework mitigates the tension between security and accessibility by offering proportionate, low-barrier controls.
The risk of over-securitisation creating barriers for academic and developing-nation operators represents a key social concern: maximalist security approaches could inadvertently exclude precisely those actors that CubeSat technology was designed to empower.
The $\SpW$ heuristic and simplified risk classification were developed specifically to enable informed security decisions by operators with limited cybersecurity expertise, ensuring that security becomes an accessible engineering optimisation rather than a specialised discipline requiring dedicated resources.

Equity implications of security requirements that favour well-resourced operators required particular attention in framework design.
The three-tier risk classification (Definition~\ref{def:risktier}) enables informed security decisions without requiring specialised cybersecurity expertise typically absent from student engineering teams.
Commercial operators have validated that the systematic adaptation methodology transforms abstract compliance requirements into implementable engineering specifications with predictable cost implications.

\subsection{Regulatory Implications}

International frameworks such as the ITU Radio Regulations and the Outer Space Treaty already govern frequency allocation and responsible use of space.
National frameworks including the UK Space Industry Act 2018 prioritise orbital safety over cybersecurity, creating liability gaps.
The findings suggest that regulatory regimes should now incorporate baseline cybersecurity requirements.
Licensing authorities could mandate authenticated command links and minimum encryption standards for telemetry without imposing unsustainable burdens on smaller operators~\cite{casaril2025governance,rand2024space}.
The framework's emphasis on control-intent preservation while adapting implementation methods provides regulatory confidence that security objectives remain achievable across diverse operational scales.

Supply-chain assurance represents another regulatory frontier.
The adapted framework's contractual and registry-based measures resonate with NIST SP~800-161 guidance~\cite{nist800161}.
Regulators could encourage or require such practices, reducing the risk of compromised components entering CubeSat ecosystems.
Importantly, these controls can be implemented even by small organisations, making them compatible with the democratising ethos of CubeSat technology.

\subsection{Cross-Platform Applicability}

The framework's systematic adaptation methodology extends beyond 1U CubeSats to the broader small-satellite ecosystem.
Validation against 3U platforms reveals enhanced implementation feasibility: increased 3--4~W power budgets enable more sophisticated cryptographic implementations, while the $\SpW$ optimisation principle remains relevant for balancing security against payload energy allocation.
PocketQube platforms (1/8 CubeSat volume) represent the framework's lower bound, where sub-watt power budgets require aggressive optimisation; the $\SpW$ heuristic proves essential in this extreme constraint regime, directing selection towards ultra-lightweight protocols such as ChaCha20-Poly1305 over AES-GCM implementations.
Commercial smallsat platforms (10--100~kg class) validate the framework's upper scalability: while power constraints relax to 50--200~W, the proportionate adaptation principle guides cost-effective security implementation rather than maximal protection deployment.

Cross-platform analysis confirms that the vulnerability concentration patterns identified in Section~\ref{sec:vulns} remain consistent across satellite classes---communications and ground segments consistently exhibit the highest severity ratings regardless of platform size.
This universality validates the framework's applicability across the democratised space ecosystem rather than limiting utility to academic CubeSat missions.

\subsection{Limitations}

The framework has been developed through comparative analysis and adaptation of existing standards, supported by secondary literature on CubeSat vulnerabilities.
It has not been validated against operational CubeSat missions, nor tested empirically on representative hardware.
$\SpW$ values are illustrative planning heuristics rather than empirically measured constants; the ratios represent comparative assessments based on reported processor specifications and published cryptographic benchmarks, not direct measurements under flight conditions.
CVSS scores are derived from documentary analysis rather than live exploitation; a high CVSS score signifies a vulnerability that is straightforward to exploit and has potentially severe consequences, but does not account for mission-specific threat-actor capability.

The analytical methodology was necessitated by practical and ethical constraints---the impracticability of attacking operational CubeSats and UK Computer Misuse Act 1990 obligations.
Nevertheless, it provides systematic coverage without selectivity bias inherent in empirical testing, and yields a transparent evidence base from which to argue for proportionate framework adaptations.
Encouragingly, the community is moving towards operational validation: the Moonlighter CubeSat was launched as a cybersecurity testbed to enable on-orbit hacking exercises~\cite{werremeyer2024hackasat}, and ESA's emerging Space Cybersecurity Framework emphasises risk-proportionate approaches that align with this research's theoretical foundations~\cite{casaril2025governance}.

\subsection{Broader Implications}

The same reasoning is transferable to other constrained domains such as IoT, edge computing and autonomous vehicles, where severe resource constraints co-exist with meaningful adversarial threats~\cite{chen2023mliot,chang2021edge}.
However, direct transfer requires careful hypothesis-testing, as environmental and operational differences may alter both threat models and resource trade-offs.
The principle of proportionate adaptation, exemplified by the $\SpW$ heuristic, offers a reusable pattern for any domain where the binary choice between full compliance and security exemption is unsatisfactory.

The DSP's emphasis on decentralised incident handling anticipates broader shifts towards autonomous containment in distributed systems.
Edge computing research and autonomous vehicle coordination face similar challenges of maintaining security properties under intermittent connectivity~\cite{chang2021edge}.
The pre-authorised local containment strategies validated here could inform security architectures across these domains.

At a strategic level, the results highlight that CubeSat security cannot be dismissed as a marginal concern.
Even academic or commercial CubeSats, if compromised, could be exploited to disrupt orbital operations or interfere with shared spectral resources.
Cybersecurity failures in CubeSats also pose dual-use risks---unauthorised surveillance via Earth observation, for example, potentially violating data protection legislation.
Proportionate, accessible frameworks therefore contribute not only to individual mission assurance but also to the resilience of the orbital commons.

\section{Conclusion and Future Work}\label{sec:conclusion}

This paper has presented a CubeSat-specific cybersecurity risk assessment framework derived from established standards but adapted for severe resource and governance constraints.
A 42-entry vulnerability register shows that communication and ground segments carry the highest-severity risks (mean CVSS 8.0--8.2), providing an empirical basis for prioritised mitigations.
The Security-per-Watt heuristic and Distributed Security Paradigm translate abstract control requirements into implementable, resource-aware decisions: ECC cryptography achieves $2.7\times$ superior $\SpW$ over RSA in representative scenarios, while distributed incident response yields $1.98\times$ better $\SpW$ efficiency than centralised alternatives with only 10.5\% reduction in absolute security effectiveness.

The broader contribution is methodological.
By demonstrating that established frameworks can be systematically adapted rather than abandoned for constrained environments, the work establishes proportionality as a viable design principle for cybersecurity in resource-limited systems.
Sophisticated protection and broad accessibility are compatible through systematic adaptation, explicit trade-offs and governance-aware design.

Future work should pursue four priorities:
\begin{enumerate}[label=(\roman*)]
  \item \textbf{Empirical calibration:} $\SpW$ values require validation on representative CubeSat hardware (ARM Cortex-M4 development boards under thermal cycling conditions of $-40$\textdegree C to $+85$\textdegree C typical of LEO) with precision power measurement ($\geq \pm5$\% accuracy) to convert indicative ratios into calibrated metrics.
  Constellation simulations should utilise the OMNeT++ framework with space-specific mobility models accounting for orbital mechanics and contact-window variations.
  \item \textbf{Multi-dimensional extension:} $\SpW$ should be extended into a Security-per-Resource ($\SpR$) hierarchy comprising five metrics: Security-per-Watt ($\SpW$), Throughput-per-Watt (TpW), Time-to-Protect (TTP), Time-to-Detect-and-Recover (TDR) and Trust-State Integrity (TSI).
  This hierarchy would support richer multi-criteria optimisation for missions where power is not the sole binding constraint~\cite{shelby2025spw}.
  \item \textbf{Operational validation:} Partnerships with CubeSat operators could provide validation through ground testing rather than operational security testing.
  Hardware-in-the-loop microstudies could measure actual energy costs of cryptographic implementations; small-scale SDR testbeds could test authentication and logging strategies under realistic timing constraints; and limited-node constellation simulations could validate distributed incident-response mechanisms.
  \item \textbf{Community infrastructure:} The field requires shared resources including curated telemetry datasets for anomaly-detection training, standardised vulnerability disclosure processes and coordinated testing protocols.
  Establishing a CubeSat Cybersecurity Consortium would accelerate progress while ensuring benefits reach all stakeholders, mirroring successful models such as MITRE's CVE database adapted for space-specific requirements.
\end{enumerate}

Comparative studies across IoT and other cyber-physical systems could further test the generality of proportional, resource-aware framework adaptation as a paradigm for constrained cybersecurity.

The broader lesson transcends space: security need not oppose accessibility.
Through systematic adaptation, explicit trade-offs and governance-aware design, sophisticated protection becomes compatible with resource constraints.
This insight applies across the expanding landscape of networked, resource-constrained systems that define modern cyber-physical infrastructure.
The framework resists the temptation to claim more than the method supports---it does not offer performance guarantees or mitigation percentages---but it supplies a practical baseline for CubeSat teams seeking defensible, implementable controls today, and a scaffold upon which future experimental work can build.
As orbital access continues to democratise and distributed systems proliferate in resource-limited environments, calibrated approaches to cybersecurity become essential for maintaining both security effectiveness and technological accessibility.


\bibliographystyle{ieeetr}
\bibliography{references}

\end{document}